\documentclass[12pt,twoside]{article}
\usepackage{graphicx}
\usepackage{hyperref} 
\begin{document}

\begin{center}
{\large\bf Quantum mechanics of bending of a nonrelativistic charged particle beam by a dipole magnet} 

\bigskip

Sameen Ahmed Khan$^a$\footnote{{\em Email}: rohelakhan@yahoo.com} and 
Ramaswamy Jagannathan$^b$\footnote{Retired Faculty, {\em  E-mail}: jagan@imsc.res.in} 

\smallskip 

$^a${\em Department of Mathematics and Sciences  \\ 
College of Arts and Applied Sciences (CAAS), Dhofar University \\ 
Post Box No. 2509, Postal Code: 211, salalah, Oman} \\ 

$^b${\em The Institute of Mathematical Sciences \\ 
4th Cross Street, Central Institutes of Technology (CIT) Campus \\ 
Tharamani, Chennai 600113, India}  
\end{center}

\begin{abstract}
Quantum mechanics of bending of a nonrelativistic monoenergetic charged particle beam by a dipole magnet is studied in the paraxial approximation.  The transfer map for the position and momentum components of a particle of the beam between two transverse planes at different points on the curved optic axis of the system is derived starting with the nonrelativistic Schr\"{o}dinger equation.  It is found that the quantum transfer map contains the classical transfer map as the main part and there are tiny quantum correction terms.  The negligibly small quantum corrections explain the remarkable success of classical mechanics in charged particle beam optics. 	
\end{abstract}
   
\medskip 
 
\noindent 
{\em Keywords}: Classical charged particle beam optics, Magnetic dipole, Bending magnet, Transfer map, Quantum Mechanics, Nonrelativistic Schr\"{o}dinger equation, Feshbach-Villars representation, Foldy-Wouthuysen transformation, Quantum charged particle beam optics, Quantum corrections. 

\setcounter{section}{0}

\section{Introduction} 
Bending, or deflection, systems are important components of many charged particle beam optical devices. 
When a particle of charge $q$ and rest mass $m$ is moving with speed $v$ along a straight line in a horizontal plane, to bend its path in a circular arc of radius $\rho$ in the horizontal plane the   constant magnetic field $B$ to be applied in the vertical direction is given by the relation   
\begin{equation}  
\frac{\gamma mv^2}{\rho} = qvB, 
\end{equation} 
so that the magnetic Lorentz force provides the required centripetal force, where 
$\gamma = 1/\sqrt{1-\left(v^2/c^2\right)}$ with $c$ as the speed of light.  In other words, 
\begin{equation}
B\rho = \frac{\gamma mv}{q} = \frac{p}{q},     
\end{equation} 
where $p$ is the momentum of the particle.  Thus, if the path of a particle of charge $q$ and momentum $p$ is to be bent along a circular arc of radius $\rho$ by a dipole magnet the constant magnetic field in the vertical direction obtained in the gap between the two poles of the magnet should be such that 
\begin{equation}
qB = \kappa p,   
\label{qbkp}
\end{equation}
where $\kappa = 1/\rho$ is the curvature of the circular orbit. Note that this formula is applicable irrespective of whether the beam is nonrelativistic ($\gamma \approx 1$) or relativistic 
($\gamma \gg 1$).    

Let us now consider a dipole magnet bending system with a circular arc of curvature $\kappa$, as the optic axis, or the design trajectory, and a magnetic field 
\begin{equation}
B_0 = \frac{\kappa p_0}{q}, 
\end{equation}
matched to the design momentum $p_0$.  The reference particle moving along the design trajectory will have $p_0$ as the longitudinal component of momentum at each point of the optic axis and zero transverse momentum components. The arclength $s$ measured along the optic axis from some reference point is the natural choice for the independent coordinate for studying the motion of the charged particle.  Let the reference particle carry an orthonormal $(x,y)$-coordinate frame with it. The $x$-axis is taken to be perpendicular to the tangent to the design orbit and in the same horizontal plane as the trajectory. The $y$-axis is taken to be in the vertical direction perpendicular to both the $x$-axis and the trajectory. The curved $s$-axis is along the design trajectory and perpendicular to both the $x$ and $y$ axes at any point on the design trajectory.  The instantaneous position of the reference particle in the design trajectory at an arc length $s$ from the reference point corresponds to $x = 0$ and $y = 0$.  Let any  particle of the beam have coordinates $(X,Y,Z)$ with respect to a fixed right handed Cartesian coordinate frame with its origin at the reference point on the design trajectory from which the arclength $s$ is measured.  Then, the two sets of coordinates of any particle of the beam, $(X,Y,Z)$ and $(x,y,s)$, are related as:  
\begin{equation}
X = \rho(1+\kappa x)\cos(\kappa s) - \rho, 
\quad Y = y, 
\quad Z = \rho(1+\kappa x)\sin(\kappa s).   
\label{QMXYZxys}
\end{equation}  

In classical charged particle beam optics, Newtonian mechanics, or the equivalent Hamiltonian mechanics, is used to study the trajectory of the charged particle moving under the Lorentz force. 
Theoretical, practical, and historical aspects of classical mechanics of electron optics have been presented extensively, with detailed references to the literature, in the first two volumes of the encyclopedic text book of Hawkes and Kasper \cite{Hawkes1, Hawkes2} and the third volume \cite{Hawkes3} presents the electron wave optics, or essentially, the quantum mechanics of electron optical imaging (for classical electron optics see also, {\em e.g.}, \cite{Wollnik}-\cite{Rose}).  In the context of accelerator physics, detailed accounts of classical charged particle beam optics are available in several books (see, {\em e.g.}, \cite{Forest}-\cite{Berz}).  In electron wave optics used for studying imaging in electron microscopy nonrelativistic Schr\"{o}dinger equation is used, and in relativistic situations either a Schr\"{o}dinger equation with relativistic correction for mass or an approximate scalar wave equation derived from the relativistic Dirac equation is used (see \cite{Hawkes3} for details).  Wondering how classical mechanics is so successful in the design and operation of electron beam optical devices like electron microscopes when the microscopic electron  should be obeying quantum mechanics, a systematic study of quantum mechanics of electron beam optics  based on the Dirac equation, the proper equation for the electron, was initiated in \cite{Jagannathan1} 
and the first quantum mechanical derivation of the classical Busch formula for the focal length of a round magnetic electron lens was obtained.  Subsequent studies have led to the formalism of quantum charged particle beam optics applicable to devices from low energy electron microscopes to high energy particle accelerators (see \cite{Jagannathan2}-\cite{Jagannathan7}). 

In \cite{Jagannathan7}, which presents a consolidated account of the formalism of quantum mechanics of charged particle beam optics, we have treated the dipole magnet, besides other charged particle beam optical elements like round magnetic lenses and magnetic quadrupoles, using the relativistic wave equations, namely, the  Klein-Gordan equation for spin-$0$ particle and the Dirac equation for spin-$\frac{1}{2}$ particle.  The Klein-Gordan equation can also be used for a particle when its spin is ignored.  We have shown, in general, how the relativistic quantum theory of any charged particle beam optical system can be approximated leading to the nonrelativistic quantum theory of the system and we have not given the details for any specific system.  This article shows explicitly how the bending of a nonrelativistic paraxial charged particle beam by a dipole magnet can be understood based on the nonrelativistic Schr\"{o}dinger equation.   

\section{Nonrelativistic classical mechanics of the optics of dipole magnet} 
Let us consider a nonrelativistic paraxial monoenergetic beam propagating through the dipole magnet.  Let $s_i$ (for $s_{in}$) and $s_o$ (for $s_{out}$) be, respectively, the entry and exit points, the points on the curved optic axis of the dipole magnet at which the beam particle enters from the free space and exits into the free space.  Classical charged particle optical Hamiltonian for the system is given by 
\begin{equation}
\mathcal{H}_o = -p_0+\frac{1}{2p_0}p_\perp^2+\frac{1}{2}p_0\kappa^2x^2,    
\label{CMdipoleH}
\end{equation} 
where $p_\perp^2 = p_x^2+p_y^2$, $p_0$ is the design momentum, and the design energy is 
$E_0 = p_0^2/2m$.  We shall derive this Hamiltonian in the next section as the classical approximation of the corresponding quantum Hamiltonian.  Hamilton's equations of motion along the $s$-axis are 
\begin{eqnarray}
\frac{dx}{ds} 
   & = & \frac{\partial\mathcal{H}_o}{\partial p_x} = \frac{p_x}{p_0},  \qquad 
\frac{dp_x}{ds} 
     = -\frac{\partial\mathcal{H}_o}{\partial x} = -p_0\kappa^2 x,  \nonumber \\   
\frac{dy}{ds} 
   & = & \frac{\partial\mathcal{H}_o}{\partial p_y} = \frac{p_y}{p_0},  \qquad 
\frac{dp_y}{ds} 
     = -\frac{\partial\mathcal{H}_o}{\partial y} = 0,   
\end{eqnarray}
leading to the paraxial equations of motion 
\begin{equation}
x^{\prime\prime}+\kappa^2x = 0, \qquad y^{\prime\prime} = 0,  
\end{equation}
where $x^{\prime\prime} = d^2x/ds^2$ and $y^{\prime\prime} = d^2y/ds^2$.  The general solutions for $x$, $x^\prime = dx/ds = p_x/p_0$, $y$ and $y^\prime = dy/ds = p_y/p_0$ at any point $s$ on the optic axis, with 
$s_i \leq s \leq s_o$, are given by 
\begin{eqnarray}
\left(\begin{array}{c}
      x(s) \\ x^\prime(s) \\ y(s) \\ y^\prime(s) 
      \end{array}\right) 
   & = & \left(\begin{array}{cccc} 
               \cos(\kappa\left(s-s_i\right)) & \frac{1}{\kappa}\sin(\kappa\left(s-s_i\right)) 
               & 0 & 0 \\
              -\kappa\sin(\kappa\left(s-s_i\right)) & \cos(\kappa\left(s-s_i\right)) & 0 & 0 \\
               0 & 0 & 1 & s-s_i \\ 0 & 0 & 0 & 1 
               \end{array}\right)  \nonumber \\   
   &   & \qquad \qquad \times \left(\begin{array}{c}
                                    x\left(s_i\right) \\ x^\prime\left(s_i\right) \\ 
                                    y\left(s_i\right) \\ y^\prime\left(s_i\right) 
                                    \end{array}\right).      
\label{paraxsoln}
\end{eqnarray} 
Note that the position coordinates of the particle $(x(s),y(s))$, and the components of the tangent to the trajectory $\left(x^\prime(s),y^\prime(s)\right)$, are the coordinates of the charged particle ray at the point $s$.  Thus the transfer map for the ray coordinates from $s_i$ to any $s$ on the optic axis is given by Eq.(\ref{paraxsoln}).  The transfer map in Eq.(\ref{paraxsoln}) shows that if a particle of the paraxial beam enters the dipole from free space along the optic axis {\em i.e.}, with $x\left(s_i\right) = 0$, $x^\prime\left(s_i\right) = 0$, $y\left(s_i\right) = 0$, and $y^\prime\left(s_i\right) = 0$, then at any point $s$ on the optic axis it will have  $x(s) = 0$, $x^\prime(s) = 0$, $y(s) = 0$, and $y^\prime(s) = 0$, or in other words, it will continue to move along the optic axis, or the design trajectory, until it exits the dipole into the free space.

It is instructive to arrive at the transfer map in Eq.(\ref{paraxsoln}) using the Poisson bracket formalism of Hamilton's equations.  Let us define the Poisson bracket between any two functions $A\left(\mathbf{r}_\perp,\mathbf{p}_\perp,s\right)$ and $B\left(\mathbf{r}_\perp,\mathbf{p}_\perp,s\right)$ as 
\begin{equation}
:A:B = \left\{A,B\right\} 
   = \left(\frac{\partial A}{\partial x}\frac{\partial B}{\partial p_x}
    -\frac{\partial A}{\partial p_x}\frac{\partial B}{\partial x}\right)
    +\left(\frac{\partial A}{\partial y}\frac{\partial B}{\partial p_y}
    -\frac{\partial A}{\partial p_y}\frac{\partial B}{\partial y}\right),   
\end{equation} 
Hamilton's equation for $s$-evolution of any observable of the particle, without explicit $s$-dependence, say $O\left(\mathbf{r}_\perp,\mathbf{p}_\perp\right)$, is 
\begin{equation}
\frac{dO}{ds} = :-\mathcal{H}_o:O.  
\end{equation} 
Since $\mathcal{H}_o$ is independent of $s$ it is possible to integrate this equation and write 
\begin{eqnarray}
O(s) & = & \left. e^{\left(s-s_i\right):-\mathcal{H}_o:}O\right|_{s=s_i}  \nonumber \\ 
     & = & \left(1+\left(s-s_i\right):-\mathcal{H}_o: 
          +\frac{\left(s-s_i\right)^2}{2!}:-\mathcal{H}_o:^2 \right. \nonumber \\ 
     &   & \qquad \qquad \left. \left.
          +\frac{\left(s-s_i\right)^3}{3!}:-\mathcal{H}_o:^3+\cdots\right)O\right|_{s=s_i}        
           \nonumber \\ 
     & = & \left(O+\left(s-s_i\right)\left\{-\mathcal{H}_o,O\right\}
          +\frac{\left(s-s_i\right)^2}{2!}
           \left\{-\mathcal{H}_o,\left\{-\mathcal{H}_o,O\right\}\right\}\right.  
           \nonumber \\
     &   & \qquad \qquad \left.\left. 
          +\frac{\left(s-s_i\right)^3}{3!}\left\{-\mathcal{H}_o,\left\{-\mathcal{H}_o,
           \left\{-\mathcal{H}_o,O\right\}\right\}\right\}+\cdots\right)\right|_{s=s_i}.          
\label{Osexp}
\end{eqnarray} 
With 
\begin{equation}
\left\{-\mathcal{H}_o,x\right\} = \frac{p_x}{p_0}, \qquad 
\left\{-\mathcal{H}_o,\frac{p_x}{p_0}\right\} = -\kappa^2 x, 
\label{HxHp}
\end{equation} 
we get 
\begin{eqnarray}
x(s) & = & \left(1-\frac{\kappa^2\left(s-s_i\right)^2}{2!}
          +\cdots\right)x\left(s_i\right)  \nonumber \\ 
     &   & \qquad +\frac{1}{\kappa}\left(\kappa\left(s-s_i\right)
          -\frac{\kappa^3\left(s-s_i\right)^3}{3!}
          +\cdots\right)\frac{p_x\left(s_i\right)}{p_0}  \nonumber \\ 
     & = & \cos(\kappa\left(s-s_i\right))x\left(s_i\right)
          +\frac{1}{\kappa}\sin(\kappa\left(s-s_i\right))\frac{p_x\left(s_i\right)}{p_0},  \nonumber \\ 
\frac{p_x(s)}{p_0} 
     & = & -\kappa\left(\kappa\left(s-s_i\right)
           -\frac{\kappa^3\left(s-s_i\right)^3}{3!}
           +\cdots\right)x\left(s_i\right)  \nonumber \\ 
     &   & \qquad +\left(1-\frac{\kappa^2\left(s-s_i\right)^2}{2!}
           +\cdots\right)\frac{p_x\left(s_i\right)}{p_0}  \nonumber \\ 
     & = & -\kappa\sin(\kappa\left(s-s_i\right))x\left(s_i\right)
           +\cos(\kappa\left(s-s_i\right))\frac{p_x\left(s_i\right)}{p_0}.    
\end{eqnarray}
With 
\begin{equation}
\left\{-\mathcal{H}_o,y\right\} = \frac{p_y}{p_0}, \qquad 
\left\{-\mathcal{H}_o,\frac{p_y}{p_0}\right\} = 0, 
\label{HyHp}
\end{equation} 
we get 
\begin{equation}
y(s) = y\left(s_i\right)+\left(s-s_i\right)\frac{p_y\left(s_i\right)}{p_0}, \qquad 
\frac{p_y(s)}{p_0} = \frac{p_y\left(s_i\right)}{p_0}.  
\end{equation}
Thus, we have the transfer map 
\begin{eqnarray}
\left(\begin{array}{c}
      x(s) \\ \frac{p_x(s)}{p_0} \\ y(s) \\ \frac{p_y(s)}{p_0} 
      \end{array}\right) 
   & = & \left(\begin{array}{cccc} 
         \cos(\kappa\left(s-s_i\right)) & \frac{1}{\kappa}\sin(\kappa\left(s-s_i\right)) 
         & 0 & 0 \\
        -\kappa\sin(\kappa\left(s-s_i\right)) & \cos(\kappa\left(s-s_i\right)) & 0 & 0 \\
         0 & 0 & 1 & \left(s-s_i\right) \\ 0 & 0 & 0 & 1 
         \end{array}\right)  \nonumber \\ 
   &   & \qquad \qquad \times \left(\begin{array}{c}
         x\left(s_i\right) \\ \frac{p_x\left(s_i\right)}{p_0} \\ 
         y\left(s_i\right) \\ \frac{p_y\left(s_i\right)}{p_0} 
         \end{array}\right), 
\label{paraxmap} 
\end{eqnarray} 
same as in Eq.(\ref{paraxsoln}) with $x^\prime(s) = p_x(s)/p_0$ and $y^\prime(s) = p_y(s)/p_0$.  For any function $A\left(\mathbf{r}_\perp,\mathbf{p}_\perp,s\right)$ the operator 
\begin{equation}
:A: = \left(\frac{\partial A}{\partial x}\frac{\partial}{\partial p_x}
     -\frac{\partial A}{\partial p_x}\frac{\partial}{\partial x}\right)
     +\left(\frac{\partial A}{\partial y}\frac{\partial}{\partial p_y}
     -\frac{\partial A}{\partial p_y}\frac{\partial}{\partial y}\right), 
\end{equation} 
such that $:A:$ acting on another function $B\left(\mathbf{r}_\perp,\mathbf{p}_\perp,s\right)$ gives the Poisson bracket $\{A,B\}$, is known as a Lie operator.  Equation (\ref{Osexp}) is a simple example of powerful Lie algebraic techniques developed to study photon optical and charged particle optical systems (see \cite{Dragt1}-\cite{Dragt4}, and also {\em e.g.}, 
\cite{Forest, Wolski, Jagannathan7, Radlicka}).  

\section{Nonrelativistic quantum mechanics of the optics of dipole magnet} 
To understand the quantum mechanics of bending of a nonrelativistic charged particle beam by a dipole magnet we have to change the corresponding Schr\"{o}dinger equation to the curved $(x,y,s)$-coordinate system and rewrite it as an $s$-evolution equation instead of time-evolution equation.  This will lead to the quantum beam optical Hamiltonian governing the $s$-evolution of the beam variables.  To this end, we proceed as follows.  
 
The free particle Schr\"{o}dinger equation in the Cartesian $(X,Y,Z)$ coordinate system  is 
\begin{eqnarray}
i\hbar\frac{\partial\Psi\left(\mathbf{R},t\right)}{\partial t} 
   & = & \widehat{H}\Psi\left(\mathbf{R},t\right)  \nonumber \\ 
\widehat{H} 
   & = & \frac{1}{2m}\widehat{P}^2 
     =   \frac{1}{2m}\left(\widehat{P}_X^2+\widehat{P}_Y^2+\widehat{P}_Z^2\right)  \nonumber \\ 
   & = & \frac{1}{2m}\left[\left(-i\hbar\frac{\partial}{\partial X}\right)^2 
                          +\left(-i\hbar\frac{\partial}{\partial Y}\right)^2
                          +\left(-i\hbar\frac{\partial}{\partial Z}\right)^2\right]  \nonumber \\ 
   & = & \frac{1}{2m}\left(-\hbar^2\nabla_{\mathbf{R}}^2\right),  
\label{Schroedinger-free}
\end{eqnarray} 
where $\widehat{H}$ is the nonrelativistic free particle Hamiltonian.  The line element $dS$, distance between two infinitesimally close points $(X,Y,Z)$ and $(X+dX,Y+dY,Z+dZ)$, corresponding to $(x,y,s)$ and $(x+dx,y+dy,s+ds)$, respectively, is given by 
\begin{equation}
dS^2 = dX^2+dY^2+dZ^2 = dx^2+dy^2+\zeta^2ds^2,
\end{equation} 
where $\zeta = 1+\kappa x$ as follows from Eq.(\ref{QMXYZxys}).  Then, we know (see, {\em e.g.}, \cite{Arfken}) that the Laplacian $\nabla_{\mathbf{R}}^2$ in the $(x,y,s)$-coordinate system becomes   
\begin{equation}
\nabla_{\mathbf{R}}^2 
   = \frac{1}{\zeta}\frac{\partial}{\partial x}\left(\zeta\frac{\partial}{\partial x}\right) 
     +\frac{\partial^2}{\partial y^2}+\frac{1}{\zeta^2}\frac{\partial^2}{\partial s^2}  
   = \frac{\partial^2}{\partial x^2}+\frac{\partial^2}{\partial y^2}
     +\frac{1}{\zeta^2}\frac{\partial^2}{\partial s^2}
     +\frac{\kappa}{\zeta}\frac{\partial}{\partial x}.  
\label{laplacian}
\end{equation} 
If we substitute this expression for $\nabla_{\mathbf{R}}^2$ in Eq.(\ref{Schroedinger-free}), and change $\Psi(\mathbf{R},t)$ to $\Psi\left(\mathbf{r}_\perp,s,t\right)$, with 
$\mathbf{r}_\perp = (x,y)$, we get 
\begin{equation}
i\hbar\frac{\partial\Psi\left(\mathbf{r}_\perp,s,t\right)}{\partial t} 
   = -\frac{\hbar^2}{2m} \left[\frac{\partial^2}{\partial x^2}+\frac{\partial^2}{\partial y^2} 
     +\frac{1}{\zeta^2}\frac{\partial^2}{\partial s^2}           
     +\frac{\kappa}{\zeta}\frac{\partial}{\partial x}\right]\Psi\left(\mathbf{r}_\perp,s,t\right).     
\label{Schrodingerxys}
\end{equation}
Note that the transformation of Eq.(\ref{Schroedinger-free}) from Cartesian coordinates to curved coordinates has resulted in Eq.(\ref{Schrodingerxys}) with a non-Hermitian Hamiltonian operator on the right hand side because of the presence of the last term, $(\kappa/\zeta)\partial/\partial x$, in the expression for $\nabla_{\mathbf{R}}^2$ in Eq.(\ref{laplacian}).  So, let us replace this term by the Hermitian term 
\begin{equation}
\frac{1}{2}\left[\left(\frac{\kappa}{\zeta}\frac{\partial}{\partial x}\right) +\left(\frac{\kappa}{\zeta} \frac{\partial}{\partial x}\right)^\dagger\right]  
   = \frac{1}{2}\left(\frac{\kappa}{\zeta}\frac{\partial}{\partial x}
    -\frac{\partial}{\partial x}\frac{\kappa}{\zeta}\right) 
   = \frac{1}{2}\left[\frac{\kappa}{\zeta}\,,\,\frac{\partial}{\partial x}\right]
   = \frac{\kappa^2}{2\zeta^2}.    
\end{equation}
Note, that $\zeta$ is a function of $x$.  Consequently, the free particle Schr\"{o}dinger equation in the curved $(x,y,s)$ coordinate system becomes 
\begin{eqnarray}
i\hbar\frac{\partial\Psi\left(\mathbf{r}_\perp,s,t\right)}{\partial t} 
   & = & \widehat{\widetilde{H}}\Psi\left(\mathbf{r}_\perp,s,t\right)  \nonumber \\  
\widehat{\widetilde{H}} 
   & = & -\frac{\hbar^2}{2m}\left[\frac{\partial^2}{\partial x^2} + \frac{\partial^2}{\partial y^2}  
         +\frac{1}{\zeta^2}\frac{\partial^2}{\partial s^2} + \frac{\kappa^2}{2 \zeta^2}\right],    
\label{schoedinger-free-curved}
\end{eqnarray} 
Let us now rewrite Eq.(\ref{schoedinger-free-curved}) as 
\begin{eqnarray}
\left(i\hbar\frac{\partial}{\partial t} \right)\Psi\left(\mathbf{r}_\perp,s,t\right) 
   & = & \frac{1}{2m}\left[\left(-i\hbar\frac{\partial}{\partial x}\right)^2 
         +\left(-i\hbar\frac{\partial}{\partial y}\right)^2 \right. \nonumber \\ 
   &   & \qquad\left. +\frac{1}{\zeta^2}\left(-i\hbar\frac{\partial}{\partial s}\right)^2 
        -\frac{\hbar^2\kappa^2}{2\zeta^2}\right]\Psi\left(\mathbf{r}_\perp,s,t\right). 
\end{eqnarray} 
To proceed from the free particle Schr\"{o}dinger equation to the equation in the presence of a  magnetic field, we can make use of the principle of electromagnetic minimal coupling 
\begin{eqnarray}
-i\hbar\frac{\partial}{\partial x} 
   & \longrightarrow & -i\hbar\frac{\partial}{\partial x}-qA_x, \quad 
-i\hbar\frac{\partial}{\partial y} 
     \longrightarrow   -i\hbar\frac{\partial}{\partial y}-qA_y,  \nonumber \\ 
\frac{1}{\zeta}\left(-i\hbar\frac{\partial}{\partial s}\right) 
   & \longrightarrow & \frac{1}{\zeta}\left(-i\hbar\frac{\partial}{\partial s}\right)-qA_s  \quad  
     = \frac{1}{\zeta}\left(-i\hbar\frac{\partial}{\partial s}-q\zeta A_s\right).                       
\label{minimal-coupling}
\end{eqnarray} 
Here $\left(A_x,A_y,A_s\right)$ are the components of the vector potential such that the magnetic field is given by 
\begin{eqnarray}
B_x & = & (\mathbf{\nabla}\times\mathbf{A})_x 
      = \frac{1}{\zeta}\left(\frac{\partial\left(\zeta A_s\right)}{\partial y}
        -\frac{\partial A_y}{\partial s}\right),  \nonumber \\ 
B_y & = & (\mathbf{\nabla}\times\mathbf{A})_y  
      = \frac{1}{\zeta}\left(\frac{\partial A_x}{\partial s}
       -\frac{\partial\left(\zeta A_s\right)}{\partial x}\right),  \nonumber \\ 
B_s & = & (\mathbf{\nabla}\times\mathbf{A})_s 
      = \left(\frac{\partial A_y}{\partial x}-\frac{\partial A_x}{\partial y}\right),   
\label{Bxys}
\end{eqnarray} 
using the expression for curl of a vector in a curved coordinate system (see, {\em e.g.}, \cite{Arfken}).  Then, the Schr\"{o}dinger equation in presence of a magnetic field, in the curved $(x,y,s)$ coordinate system, becomes 
\begin{eqnarray}
i\hbar\frac{\partial\Psi\left(\mathbf{r}_\perp,s,t\right)}{\partial t}       
   & = & \widehat{\mathcal{H}}\Psi\left(\mathbf{r}_\perp,s,t\right),  \nonumber \\ 
\widehat{\mathcal{H}}
   & = & \frac{1}{2m}\left[\left(\widehat{\pi}_x^2+\widehat{\pi}_y^2+\widehat{\pi}_s^2\right) 
        -\frac{\hbar^2\kappa^2}{2\zeta^2}\right],  
\end{eqnarray} 
with 
\begin{eqnarray}
\widehat{\pi}_x 
   & = & \widehat{p}_x-qA_x = \left(-i\hbar\frac{\partial}{\partial x}-qA_x\right),  \nonumber \\  
\widehat{\pi}_y 
   & = & \widehat{p}_y-qA_y = \left(-i\hbar\frac{\partial}{\partial y}-qA_y\right),  \nonumber \\ 
\widehat{\pi}_s 
   & = & \widehat{p}_s-qA_s = \left(\frac{-i\hbar}{\zeta}\frac{\partial}{\partial s}-qA_s\right).     
\end{eqnarray}
Our aim is to study the evolution of the beam characteristics along the optic axis of the system. For  this we have to rewrite the Schr\"{o}dinger equation as an $s$-evolution equation.  To this end we proceed as follows.  

Let us consider a nonrelativistic monoenergetic paraxial charged particle beam propagating through the dipole magnet.  We can associate the beam particle with the wave function 
\begin{equation}  
\Psi\left(\mathbf{r}_\perp,s,t\right) = e^{-iE_0t/\hbar}\psi\left(\mathbf{r}_\perp,s\right), \quad 
\mbox{with}\ \ E_0 = \frac{p_0^2}{2m}, 
\end{equation} 
corresponding to a scattering state, where $p_0$ is the design momentum with which the particle enters the bending magnet from the free space outside.  Then, $\psi\left(\mathbf{r}_\perp,s\right)$ satisfies the time-independent Schr\"{o}dinger equation 
\begin{equation} 
\widehat{\mathcal{H}}\psi\left(\mathbf{r}_\perp,s\right) 
   = \frac{1}{2m}\left[\left(\widehat{\pi}_x^2+\widehat{\pi}_y^2+\widehat{\pi}_s^2\right) 
    -\frac{\hbar^2\kappa^2}{2\zeta^2}\right]\psi\left(\mathbf{r}_\perp,s\right) 
   = E_0\psi\left(\mathbf{r}_\perp,s\right).  
\label{QtiSEqpi}
\end{equation} 
Rearranging the terms in Eq.(\ref{QtiSEqpi}), we can rewrite it as 
\begin{eqnarray}
\widehat{\pi}_s^2\psi\left(\mathbf{r}_\perp,s\right) 
   & = & \left(2mE_0-\widehat{\pi}_x^2-\widehat{\pi}_y^2+\frac{\hbar^2\kappa^2}{2\zeta^2}\right)
         \psi\left(\mathbf{r}_\perp,s\right) \nonumber \\            
   & = & \left(p_0^2-\widehat{\pi}_x^2-\widehat{\pi}_y^2+\mathsf{p}^2\right)
         \psi\left(\mathbf{r}_\perp,s\right), 
\label{pissqreq}
\end{eqnarray} 
where 
\begin{equation}  
p_0^2 = 2mE_0, \qquad 
\mathsf{p}^2 = \frac{\hbar^2\kappa^2}{2\zeta^2}.  
\end{equation} 
For the dipole magnet we can take  
\begin{equation} 
A_x = 0, \quad 
A_y = 0, \quad 
A_s = -B_0\left(x-\frac{\kappa x^2}{2\zeta}\right), 
\end{equation} 
so that the magnetic field is given by 
\begin{equation}
B_x = 0, \quad 
B_y = B_0, \quad 
B_s = 0,   
\end{equation} 
according to Eq.(\ref{Bxys}).  Then, we have 
\begin{equation}
\widehat{\pi}_s^2\psi\left(\mathbf{r}_\perp,s\right) 
   = \left(p_0^2 - \widehat{p}_\perp^2 + \mathsf{p}^2\right)\psi\left(\mathbf{r}_\perp,s\right),   
\label{pissqreq2}
\end{equation}  
where $\widehat{p}_\perp^2 = \widehat{p}_x^2+\widehat{p}_y^2$.

Following the idea of the Feshbach-Villars representation of the Klein-Gordon equation (see, 
{\em e.g.}, \cite{Bjorken, Greiner1}, and also \cite{Jagannathan7, Khan7, Khan8}), let us now rewrite Eq.(\ref{pissqreq2}) as an equation for a two-component wave function.  First, we write it as  
\begin{equation}
\frac{\widehat{\pi}_s}{p_0}\left(\begin{array}{c}
                                 \psi_1 \\ \psi_2  
                                 \end{array}\right) 
   = \left(\begin{array}{cc}
           0 & 1 \\ 
           \frac{1}{p_0^2}\left(p_0^2-\widehat{p}_\perp^2+\mathsf{p}^2\right) & 0 
           \end{array}\right)
     \left(\begin{array}{c}
           \psi_1 \\ \psi_2  
           \end{array}\right),    
\end{equation}
with 
\begin{equation}
\left(\begin{array}{c}
      \psi_1 \\ \psi_2  
      \end{array}\right) 
   = \left(\begin{array}{c}
     \psi \\ \frac{\widehat{\pi_s}}{p_0}\psi 
   \end{array}\right).  
\end{equation}
Now, let 
\begin{equation}
\left(\begin{array}{c}
      \psi_+ \\ \psi_- 
      \end{array}\right) 
   = \frac{1}{2}\left(\begin{array}{c}
                      \psi_1 + \psi_2 \\ \psi_1 - \psi_2
                      \end{array}\right)          
   = \frac{1}{2}\left(\begin{array}{cr}
                      1 & 1 \\ 1 & -1
                      \end{array}\right)
                \left(\begin{array}{c}
                      \psi_1 \\ \psi_2  
                      \end{array}\right)    
   = M\left(\begin{array}{c}
            \psi_1 \\ \psi_2  
            \end{array}\right).  
\end{equation} 
Then, it follows that 
\begin{eqnarray}
\frac{\widehat{\pi}_s}{p_0}\left(\begin{array}{c}
                                 \psi_+ \\ \psi_- 
                                 \end{array}\right) 
   & = & \frac{1}{p_0}\left(\frac{-i\hbar}{\zeta}
                            \frac{\partial}{\partial s}-qA_s\right) 
                      \left(\begin{array}{c}         
                            \psi_+ \\ \psi_- 
                            \end{array}\right)  \nonumber \\ 
   & = & M\left(\begin{array}{cc}
                0 & 1 \\ 
                \frac{1}{p_0^2}\left(p_0^2-\widehat{p}_\perp^2+\mathsf{p}^2\right) & 0 
                \end{array}\right)M^{-1}
          \left(\begin{array}{c}
                \psi_+ \\ \psi_-                             
                \end{array}\right)  \nonumber \\ 
   & = & \left(\begin{array}{cc}
               1-\frac{1}{2p_0^2}\left(\widehat{p}_\perp^2-\mathsf{p}^2\right) & 
                -\frac{1}{2p_0^2}\left(\widehat{p}_\perp^2-\mathsf{p}^2\right) \\ 
                 \frac{1}{2p_0^2}\left(\widehat{p}_\perp^2-\mathsf{p}^2\right) & 
              -1+\frac{1}{2p_0^2}\left(\widehat{p}_\perp^2-\mathsf{p}^2\right)  
               \end{array}\right)
         \left(\begin{array}{c}
               \psi_+ \\ \psi_- 
               \end{array}\right). \nonumber\\ 
   &   &  
\end{eqnarray} 
Rearranging this equation we get  
\begin{eqnarray} 
i\hbar\frac{\partial}{\partial s}\left(\begin{array}{c}
                                       \psi_+ \\ \psi_- 
                                       \end{array}\right)
   & = & \widehat{\mathtt{H}}\left(\begin{array}{c}
                                   \psi_+ \\ \psi_- 
                                   \end{array}\right),  \nonumber \\ 
\widehat{\mathtt{H}} 
   & = & -p_0\sigma_z + \widehat{\mathcal{E}} + \widehat{\mathcal{O}},  \nonumber \\ 
\widehat{\mathcal{E}} 
   & = & -q\zeta A_sI+\left(\frac{\zeta}{2p_0}\left(\widehat{p}_\perp^2
         -\mathsf{p}^2\right)-p_0\kappa x\right)\sigma_z,  \nonumber \\ 
\widehat{\mathcal{O}} 
   & = & \frac{i\zeta}{2p_0}\left(\widehat{p}_\perp^2-\mathsf{p}^2\right)\sigma_y.   
\label{mathttH}
\end{eqnarray}  
Let us consider a particle of the paraxial beam in the free space outside the dipole just before  entering the dipole.  For this particle $s$-axis becomes $z$-axis and its longitudinal momentum $p_z$ will be $< p_0$ and $\approx p_0$.  Its transverse momentum components will be $\ll p_0$ and 
$\approx 0$. In other words, for this particle $p_z/p_0 \approx 1$.  For this free particle entering the dipole we can take the wave function to be of the form 
$\psi = \phi\left(\mathbf{r}_\perp\right)e^{ip_zz/\hbar}$ and correspondingly we have 
$\psi_+ = \frac{1}{2}\left[\psi+\left(\widehat{p}_z/p_0\right)\psi\right] \approx \psi$ and 
$\psi_- = \frac{1}{2}\left[\psi-\left(\widehat{p}_z/p_0\right)\psi\right] \approx 0$.  This shows that for a particle of a forward moving paraxial beam $\psi_+$ will be large compared to $\psi_-$.  

In Eq.(\ref{mathttH}), $I$ is the $2\times 2$ identity matrix and 
\begin{equation} 
\sigma_z = \left(\begin{array}{cr}
                 1 & 0 \\ 0 & -1 
                 \end{array}\right), \quad  
\sigma_y = \left(\begin{array}{cr}
                 0 & -i \\ i & 0  
                 \end{array}\right),              
\end{equation} 
are two of the well known triplet of Pauli matrices.  The $2\times 2$ matrix operator $\widehat{\mathcal{O}}$ which has nonzero entries only along the off-diagonal and couples the upper and lower components of any two-component wave function on which it acts is known as an odd term of $\widehat{\mathtt{H}}$. The $2\times 2$ matrix operator $\widehat{\mathcal{E}}$ which has nonzero entries only along the diagonal and does not couple the upper and lower components of any two-component wave function on which it acts is known as an even term.  Note that $\sigma_z$ commutes with $\widehat{\mathcal{E}}$ and anticommutes with $\widehat{\mathcal{O}}$: 
\begin{equation} 
\sigma_z\widehat{\mathcal{E}} = \widehat{\mathcal{E}}\sigma_z, \quad 
\sigma_z\widehat{\mathcal{O}} = -\widehat{\mathcal{O}}\sigma_z. 
\label{evenoddcommrel}
\end{equation}
The structure of Eq.(\ref{mathttH}) is similar to the structure of the Dirac equation in which also the  Hamiltonian can be split in to a constant term $mc^2\beta$ (like $-p_0\sigma_z$) and an even and an odd term which, respectively, commute and anticommute with $\beta$.  The Dirac wave function has four components and for a positive energy Dirac particle the upper pair of components are large compared to the lower pair of components in the nonrelativistic situation.  Based purely on the commutation relations between $\beta$ and the even and odd terms, same as Eq.(\ref{evenoddcommrel})  with $\beta$ instead of $\sigma_z$, a series of Foldy-Wouthuysen transformations (see, {\em e.g.}, \cite{Bjorken, Greiner1}, and also \cite{Jagannathan7, Khan7, Khan8}) leads to an expansion of the Dirac Hamiltonian into nonrelativistic approximation and a series of relativistic correction terms.  Following a similar Foldy-Wouthuysen-like transformation technique we can expand $\widehat{\mathtt{H}}$ into paraxial approximation and a series of nonparaxial, or aberration, correction terms.  We shall be interested only in the paraxial approximation.    

Let us now apply the first Foldy-Wouthuysen-like transformation:   
\begin{equation}  
\left(\begin{array}{c}
      \psi_+^{(1)} \\ \psi_-^{(1)} 
      \end{array}\right) 
   = e^{i\widehat{S}_1}\left(\begin{array}{c}
                             \psi_+ \\ \psi_-
                             \end{array}\right), \quad 
\mbox{with}\ \ \widehat{S}_1 = \frac{i}{2p_0}\sigma_z\widehat{\mathcal{O}}.  
\label{FWlikeT}
\end{equation} 
The result is 
\begin{eqnarray}
i\hbar\frac{\partial}{\partial s}\left(\begin{array}{c}
                                       \psi_+^{(1)} \\ \psi_-^{(1)}
                                       \end{array}\right) 
   & = & \widehat{\mathtt{H}}^{(1)}\left(\begin{array}{c}
                                         \psi_+^{(1)} \\ \psi_-^{(1)}
                                         \end{array}\right),  \nonumber \\ 
\widehat{\mathtt{H}}^{(1)}                                                                          
   & = & e^{i\widehat{S}_1}\widehat{\mathtt{H}}e^{-i\widehat{S}_1}
        -i\hbar e^{i\widehat{S}_1}\frac{\partial}{\partial s} \left(e^{-i\widehat{S}_1}\right)  
     = e^{i\widehat{S}_1}\widehat{\mathtt{H}}e^{-i\widehat{S}_1}, 
\end{eqnarray} 
since $\widehat{S}_1$ is $s$-independent.  Calculating $\widehat{\mathtt{H}}^{(1)}$, using the operator identity
\begin{equation}  
e^{\widehat{A}}\widehat{B}e^{-\widehat{A}} 
 = \widehat{B}+\left[\widehat{A},\widehat{B}\right]
   +\frac{1}{2!}\left[\widehat{A},\left[\widehat{A},\widehat{B}\right]\right]
   +\frac{1}{3!}\left[\widehat{A},\left[\widehat{A},\left[\widehat{A},\widehat{B}\right]\right]\right] 
   +\cdots, 
\label{identity}
\end{equation} 
with the commutator bracket $\left[\widehat{A},\widehat{B}\right] = \left(\widehat{A}\widehat{B}-\widehat{B}\widehat{A}\right)$, we have 
\begin{equation}
\widehat{\mathtt{H}}^{(1)}   
   \approx -q\zeta A_sI-\left[\zeta p_0
           -\frac{1}{2p_0}\left(\widehat{p}_\perp^2-\zeta\mathsf{p}^2\right)\right]\sigma_z,   
\end{equation} 
up to the paraxial approximation in which only terms upto quadratic in $\mathbf{r}_\perp$ and $\widehat{\mathbf{p}}_\perp$ are retained.  From Eq.(\ref{mathttH}) and Eq.(\ref{FWlikeT}) it can be seen that for a forward-moving paraxial beam $\psi_+^{(1)}$ is large compared to $\psi_-^{(1)}$, exactly as we found in the case of $\psi_+$ and $\psi_-$.  In view of this, we can approximate $\widehat{\mathtt{H}}^{(1)}$ further by replacing $\sigma_z$ by $I$.  Thus, we write 
\begin{equation}
\widehat{\mathtt{H}}^{(1)}    
   \approx -q\zeta A_s-\zeta p_0+\frac{1}{2p_0}\left(\widehat{p}_\perp^2-\zeta\mathsf{p}^2\right). 
\label{sigmaz2i}
\end{equation}  
We can take 
$\zeta\mathsf{p}^2 = \hbar^2\kappa^2/2\zeta \approx \frac{1}{2}\hbar^2\kappa^2(1-\kappa x)$ since $\kappa x \ll 1$ for a paraxial beam.  We are interested in the $s$-evolution of $\psi(\mathbf{r}_\perp,s)$.  To this end, we retrace the above transformations:   
\begin{eqnarray}
i\hbar\frac{\partial}{\partial s}\left(\begin{array}{c}
                                       \psi_1 \\ \psi_2 
                                       \end{array}\right) 
   & = & i\hbar\frac{\partial}{\partial s}
         \left[M^{-1}e^{-i\widehat{S}_1}\left(\begin{array}{c}
                                              \psi^{(1)}_+ \\ \psi^{(1)}_- 
                                              \end{array}\right)\right]  \nonumber \\ 
   & = & M^{-1}e^{-i\widehat{S}_1}\left[i\hbar\frac{\partial}{\partial s}
                                  \left(\begin{array}{c}
                                        \psi^{(1)}_+ \\ \psi^{(1)}_- 
                                        \end{array}\right)\right]  \nonumber \\ 
   & = & \left(M^{-1}e^{-i\widehat{S}_1}\widehat{\mathtt{H}}^{(1)}e^{i\widehat{S}_1}M\right) 
         \left(\begin{array}{c}
               \psi_1 \\ \psi_2 
               \end{array}\right).                                                          
\label{retrace}
\end{eqnarray} 
One may wonder whether such a retracing of the transformations would lead back to the original equation (\ref{mathttH}).  It is not so because of the truncation of the series with the paraxial term and the approximation of replacing $\sigma_z$ by $I$ in Eq.(\ref{sigmaz2i}).  The result of this step will be the desired paraxial approximation of the original equation (\ref{mathttH}).  Thus, from the equation  obtained from (\ref{retrace}) we can identify the $s$-evolution equation for $\psi(\mathbf{r}_\perp,s)$ since $\psi(\mathbf{r}_\perp,s) = \psi_1$.  The result is:   
\begin{eqnarray}
i\hbar\frac{\partial}{\partial s}\psi\left(\mathbf{r}_\perp,s\right) 
   & = & \left[-p_0+\frac{1}{2p_0}\widehat{p}_\perp^2+\frac{1}{2}qB_0\kappa x^2 \right. \nonumber \\  
   &   & \qquad \left. +\left(qB_0-\kappa p_0\right)x-\frac{\hbar^2\kappa^2}{4p_0}(1-\kappa x)\right]
         \psi\left(\mathbf{r}_\perp,s\right).    
\end{eqnarray} 
With the curvature of the design trajectory matched to the dipole magnetic field as   
\begin{equation}
qB_0 = \kappa p_0, 
\end{equation} 
we have 
\begin{eqnarray}
i\hbar\frac{\partial}{\partial s}\psi\left(\mathbf{r}_\perp,s\right) 
   & = & \widehat{\mathcal{H}}_o\psi\left(\mathbf{r}_\perp,s\right), 
         \nonumber \\ 
\widehat{\mathcal{H}}_o   
   & = & \left[-p_0+\frac{1}{2p_0}\widehat{p}_\perp^2+\frac{1}{2}p_0\kappa^2x^2 \right. \nonumber \\ 
   &   & \qquad \qquad \qquad \left. -\frac{\hbar^2\kappa^2}{4p_0}(1-\kappa x)\right]
         \psi\left(\mathbf{r}_\perp,s\right).  
\label{QMdipoleH}
\end{eqnarray} 
This $\widehat{\mathcal{H}}_o$ is the quantum charged particle beam optical Hamiltonian of the dipole magnet for a paraxial beam.  

The wave function $\psi\left(\mathbf{r}_\perp,s\right)$ represents the probability amplitude, and 
$\left|\psi\left(\mathbf{r}_\perp,s\right)\right|^2$ represents the probability, for the particle to be found at the position $(x,y)$ in the transverse plane at the point $s$ on the optic axis.  Since $\mathcal{H}_o$ is Hermitian, $s$-evolution of $\psi\left(\mathbf{r}_\perp,s\right)$ is unitary and the normalization 
\begin{equation} 
\int_{-\infty}^{\infty}dxdy\ \left|\psi\left(\mathbf{r}_\perp,s\right)\right|^2 
   = \left\langle\psi(s)\right|\left.\psi(s)\right\rangle = 1, 	
\end{equation} 
will be preserved.  For any observable $O\left(\mathbf{r}_\perp,\mathbf{p}_\perp\right)$ of the particle, represented by the corresponding Hermitian operator $\widehat{O}\left(\mathbf{r}_\perp,\widehat{\mathbf{p}}_\perp\right)$, we have  
\begin{equation}
\left\langle\widehat{O}\right\rangle(s) 
   = \int_{-\infty}^{\infty}dxdy\ \psi^*\left(\mathbf{r}_\perp,s\right)\widehat{O}
                                 \psi\left(\mathbf{r}_\perp,s\right)
   = \left\langle\psi(s)\right|\widehat{O}\left|\psi(s)\right\rangle 
\end{equation}
as the average value in the transverse plane at $s$.  We can take the average value  $\left\langle\widehat{O}\right\rangle(s)$ as corresponding to the value of the classical variable $O$ 
in the transverse plane at $s$, following the Ehrenfest theorem (see, {\em e.g.}, 
\cite{Greiner2, Griffiths}, and also \cite{Jagannathan7}).  We shall be interested in the $s$-evolution of $\left\langle\widehat{x}\right\rangle(s)$, $\left\langle\widehat{p}_x\right\rangle(s)/p_0$, $\left\langle\widehat{y}\right\rangle(s)$, and $\left\langle\widehat{p}_y\right\rangle(s)/p_0$.  
If we drop the last $\hbar$-dependent term from $\widehat{\mathcal{H}}_o$ and replace the quantum operators $x$ and $\widehat{p}_\perp^2$, respectively, by the corresponding classical variables $x$ and $p_\perp^2$ we get the classical charged particle beam optical Hamiltonian of the dipole magnet for a paraxial beam,  
\begin{equation}
\mathcal{H}_o = -p_0+\frac{1}{2p_0}p_\perp^2+\frac{1}{2}p_0\kappa^2x^2,    
\end{equation} 
same as taken in Eq.(\ref{CMdipoleH}).  

Since the quantum beam optical Hamiltonian $\widehat{\mathcal{H}}_o$ is independent of $s$ we can integrate Eq.(\ref{QMdipoleH}) and write 
\begin{equation}
\psi\left(\mathbf{r}_\perp,s\right) 
   = e^{-\frac{i}{\hbar}\left(s-s_i\right)\widehat{\mathcal{H}}_o}
     \psi\left(\mathbf{r}_\perp,s_i\right), 
\end{equation}
or 
\begin{equation}
|\psi(s)\rangle 
   = e^{-\frac{i}{\hbar}\left(s-s_i\right)\widehat{\mathcal{H}}_o}
     \left|\psi\left(s_i\right)\right\rangle.   
\end{equation}
Then, we can write the average value $\left\langle\widehat{O}\right\rangle(s)$ for any $\widehat{O}$ as 
\begin{equation}
\left\langle\widehat{O}\right\rangle(s) 
   = \left\langle\psi\left(s_i\right)\right|
     e^{\frac{i}{\hbar}\left(s-s_i\right)\widehat{\mathcal{H}}_o}\widehat{O}
     e^{-\frac{i}{\hbar}\left(s-s_i\right)\widehat{\mathcal{H}}_o}
     \left|\psi\left(s_i\right)\right\rangle.
\end{equation} 
Using the operator identity in Eq.(\ref{identity}), and defining 
\begin{equation} 
:\widehat{A}:\widehat{B} = \left[\widehat{A},\widehat{B}\right], 
\end{equation} 
we can write  
\begin{eqnarray} 
e^{\frac{i}{\hbar}\left(s-s_i\right)\widehat{\mathcal{H}}_o}\widehat{O}
e^{-\frac{i}{\hbar}\left(s-s_i\right)\widehat{\mathcal{H}}_o} 
   & = & \left[I+\frac{i}{\hbar}\left(s-s_i\right):\widehat{\mathcal{H}}_o: 
        +\frac{1}{2!}\left(\frac{i}{\hbar}\left(s-s_i\right)\right)^2:\widehat{\mathcal{H}}_o:^2  
         \right. \nonumber \\ 
   &   & \qquad \left.    
        +\frac{1}{3!}\left(\frac{i}{\hbar}\left(s-s_i\right)\right)^3:\widehat{\mathcal{H}}_o:^3
        +\cdots\right]\widehat{O}  \nonumber \\ 
   & = & e^{\frac{i}{\hbar}\left(s-s_i\right):\widehat{\mathcal{H}}_o:}\widehat{O}.    
\end{eqnarray}
Thus, we write 
\begin{equation}
\left\langle\widehat{O}\right\rangle(s) 
   = \left\langle\psi\left(s_i\right)\right|
     e^{\frac{i}{\hbar}\left(s-s_i\right):\widehat{\mathcal{H}}_o:}\widehat{O}
     \left|\psi\left(s_i\right)\right\rangle 
   = \left\langle e^{\frac{i}{\hbar}\left(s-s_i\right):\widehat{\mathcal{H}}_o:}\widehat{O}
     \right\rangle\left(s_i\right).  
\label{QMOsexp}
\end{equation}
It may be noted that, considering the quantum average of an observable as the value of the corresponding classical observable following the Ehrenfest theorem, the transition from the quantum transfer map in Eq.(\ref{QMOsexp}) to the classical transfer map in Eq.(\ref{Osexp}) is in accordance with the classical $\longrightarrow$ quantum correspondence rule of Dirac: 
\begin{equation} 
\{A,B\} \longrightarrow \frac{1}{i\hbar}\left[\widehat{A},\widehat{B}\right].  
\end{equation} 
Thus, in the transition from Eq.(\ref{QMOsexp}) to Eq.(\ref{Osexp}), 
$(i/\hbar)\left[\widehat{\mathcal{H}}_o\ ,\ \cdot\right] 
= (1/i\hbar)\left[-\widehat{\mathcal{H}}_o\ ,\ \cdot\right]$ gets replaced by  
$\left\{-\mathcal{H}_o\ ,\ \cdot\right\}$.  

From Eq.(\ref{QMOsexp}) it follows that   
\begin{equation}
\left(\begin{array}{c} 
      \langle x\rangle(s) \\ \frac{1}{p_0}\left\langle\widehat{p}_x\right\rangle(s) \\ 
      \langle y\rangle(s) \\ \frac{1}{p_0}\left\langle\widehat{p}_y\right\rangle(s) 
      \end{array}\right) 
   = \left(\begin{array}{c}
         \left\langle e^{\frac{i}{\hbar}\left(s-s_i\right) :\widehat{\mathcal{H}}_o:}x\right\rangle\left(s_i\right) \\
         \frac{1}{p_0}\left\langle e^{\frac{i}{\hbar}\left(s-s_i\right):\widehat{\mathcal{H}}_o:}
         \widehat{p}_x\right\rangle\left(s_i\right) \\ 
         \left\langle e^{\frac{i}{\hbar}\left(s-s_i\right) :\widehat{\mathcal{H}}_o:}y\right\rangle\left(s_i\right) \\
         \frac{1}{p_0}\left\langle e^{\frac{i}{\hbar}\left(s-s_i\right) :\widehat{\mathcal{H}}_o:}\widehat{p}_y\right\rangle\left(s_i\right) 
         \end{array}\right). 
\end{equation}  
Omitting from $\widehat{\mathcal{H}}_o$ the constant terms $-p_0$ and $-\hbar^2\kappa^2/4p_0$ which have zero commutator with the quantum operator corresponding to any observable, we have 
\begin{eqnarray}
\frac{i}{\hbar}:\widehat{\mathcal{H}}_o:x 
   & = & \frac{i}{\hbar}\left[\frac{1}{2p_0}\widehat{p}_\perp^2
         +\frac{1}{2}p_0\kappa^2x^2+\frac{\hbar^2\kappa^3}{4p_0}x\,,\,x\right] 
     =   \frac{\widehat{p}_x}{p_0},  \nonumber \\ 
\frac{i}{\hbar}:\widehat{\mathcal{H}}_o:\frac{\widehat{p}_x}{p_0}  
   & = & \frac{i}{\hbar}\left[\frac{1}{2p_0}\widehat{p}_\perp^2
         +\frac{1}{2}p_0\kappa^2x^2
         +\frac{\hbar^2\kappa^3}{4p_0}x\,,\,\frac{\widehat{p}_x}{p_0}\right]  
     =   -\kappa^2 x-\frac{\hbar^2\kappa^3}{4p_0^2},  \nonumber \\  
\frac{i}{\hbar}:\widehat{\mathcal{H}}_o:y 
   & = & \frac{i}{\hbar}\left[\frac{1}{2p_0}\widehat{p}_\perp^2
         +\frac{1}{2}p_0\kappa^2x^2+\frac{\hbar^2\kappa^3}{4p_0}x\,,\,y\right]  
     =   \frac{\widehat{p}_y}{p_0},  \nonumber \\ 
\frac{i}{\hbar}:\widehat{\mathcal{H}}_o:\frac{\widehat{p}_y}{p_0}  
   & = & \frac{i}{\hbar}\left[\frac{1}{2p_0}\widehat{p}_\perp^2
        +\frac{1}{2}p_0\kappa^2x^2
        +\frac{\hbar^2\kappa^3}{4p_0}x\,,\,\frac{\widehat{p}_y}{p_0}\right] = 0,  
\end{eqnarray}
leading to the quantum transfer map for the dipole 
\begin{eqnarray}
\left(\begin{array}{c} 
      \langle x\rangle(s) \\ \frac{1}{p_0}\left\langle\widehat{p}_x\right\rangle(s) \\ 
      \langle y\rangle(s) \\ \frac{1}{p_0}\left\langle\widehat{p}_y\right\rangle(s) 
      \end{array}\right) 
   & = & \left(\begin{array}{cccc}
               \cos(\kappa\left(s-s_i\right)) & 
               \frac{1}{\kappa}\sin(\kappa\left(s-s_i\right)) & 0 & 0 \\   
              -\kappa\sin(\kappa\left(s-s_i\right)) & 
               \cos(\kappa\left(s-s_i\right)) & 0 & 0 \\ 
               0 & 0 & 1 & s-s_i \\ 
               0 & 0 & 0 & 1  
               \end{array}\right) \nonumber \\   
   &   & \qquad\times \left(\begin{array}{c}
                \left\langle x\right\rangle\left(s_i\right) \\
                \frac{1}{p_0}\left\langle\widehat{p}_x\right\rangle\left(s_i\right) \\ 
                \left\langle y\right\rangle\left(s_i\right) \\
                \frac{1}{p_0}\left\langle\widehat{p}_y\right\rangle\left(s_i\right) 
                \end{array}\right)  \nonumber \\ 
   &   & \qquad +\left(\begin{array}{c} 
                      \frac{\hbar^2\kappa}{4p_0^2} 
                      (\cos(\kappa\left(s-s_i\right))-1) \\ 
                      \frac{\hbar^2\kappa^2}{4p_0^2}
                      \sin(\kappa\left(s-s_i\right)) \\ 0 \\ 0 
                      \end{array}\right).  
\label{QTM}
\end{eqnarray} 
This shows that a particle entering the dipole magnet along the design trajectory, {\em i.e.}, with  
$\left\langle x\right\rangle\left(s_i\right) = 0,\  
\left\langle\widehat{p}_x\right\rangle\left(s_i\right) = 0,\    
\left\langle y\right\rangle\left(s_i\right) = 0,\ 
\left\langle\widehat{p}_y\right\rangle\left(s_i\right) = 0$, will follow the curved  design trajectory, except for some tiny quantum kicks in the $x$ coordinate 
$(\sim \lambda_0^2/\rho)$ and the $x$-gradient $(\sim \lambda_0^2/\rho^2)$, where 
$\lambda_0$ is the de Broglie wavelength!  The quantum transfer map in Eq.(\ref{QTM}) is seen to contain the classical transfer map as the main part and the quantum correction part is negligibly small. This explains the remarkable effectiveness of classical charged particle beam optics in the design and operation of bending magnets.  Analysis of the quantum mechanics of some other charged particle optical elements has also led to the same conclusion (see \cite{Jagannathan7}).   

\section{Concluding Remarks}
To summarize, we have studied the bending of a nonrelativistic monoenergetic charged particle beam by a dipole magnet in the paraxial approximation using quantum mechanics.  This involves rewriting the nonrelativistic Schr\"{o}dinger equation as an equation of evolution along the curved optic axis of the system and calculating the transfer map for the quantum averages of transverse position and momentum components. Classical mechanics of the optics of dipole magnet is seen to emerge as an approximation  of the quantum theory.  It is found that a particle entering the dipole magnet along the design trajectory, will follow the curved  design trajectory, except for some tiny quantum kicks in the position $(\sim \lambda_0^2/\rho)$ and the gradient $(\sim \lambda_0^2/\rho^2)$, where $\lambda_0$ is the de Broglie wavelength!  In the relativistic case the study of quantum mechanics of the optics of the dipole magnet using the Klein-Gordon equation leads to the same transfer map (\ref{QTM}) with $p_0$  replaced by its relativistic expression $\sqrt{\left(E_0/c\right)^2-m^2c^2}$ (see \cite{Jagannathan7} for details), and consequently the quantum corrections become still smaller due to shorter de Broglie wavelengths.  Such quantum corrections could have significant effects in the case of ultracold electron beams.  The negligibility of quantum corrections explains the remarkable success of classical mechanics in the design and operation of charged particle beam optical devices from electron microscopes to particle accelerators.

\end{document}